\documentclass[conference]{IEEEtran}

\usepackage{cite}
\usepackage{amsmath,amssymb,amsfonts}
\usepackage{algorithmic}
\usepackage{graphicx}
\usepackage{textcomp}
\usepackage{xcolor}
\usepackage{booktabs}
\usepackage[hyphens]{url}
\usepackage{hyperref}
\usepackage[hyphenbreaks]{breakurl} 
\usepackage{listings}
\usepackage{mdframed}
\def\BibTeX{{\rm B\kern-.05em{\sc i\kern-.025em b}\kern-.08em
    T\kern-.1667em\lower.7ex\hbox{E}\kern-.125emX}}

\newcommand{\greybox}[1]{
    \begin{mdframed}[backgroundcolor=black!10!white,linewidth=0pt,backgroundcolor=black!10,linewidth=0pt,innerleftmargin=5pt,innertopmargin=5pt]
        #1
    \end{mdframed}
}

\newcommand{\RQOne}{What syntactic and semantic errors do students make?}
\newcommand{\RQThree}{What logical errors do students make?}

\begin{document}
\lstset{
  language=Java, 
  basicstyle=\ttfamily\footnotesize, 
  keywordstyle=\color{blue}, 
  stringstyle=\color{red}, 
  commentstyle=\color{green}, 
  numbers=left, 
  numberstyle=\tiny\color{gray}, 
  breaklines=true, 
  frame=single, 
  backgroundcolor=\color{lightgray!10}, 
  tabsize=2 
}

\title{From Bugs to Breakthroughs: Novice Errors in CS2\\
}
\author{\IEEEauthorblockN{Nadja Just}
 \IEEEauthorblockA{\textit{Professorship of Software Engineering} \\
 \textit{University of Technology Chemnitz}\\
 Chemnitz, Germany \\
 https://orcid.org/0009-0006-9076-0362}
 \and
 \IEEEauthorblockN{Belinda Schantong}
 \IEEEauthorblockA{\textit{Professorship of Software Engineering} \\
 \textit{University of Technology Chemnitz}\\
 Chemnitz, Germany \\
 https://orcid.org/0009-0005-7091-0880}
 \and
 \IEEEauthorblockN{Janet Siegmund}
 \IEEEauthorblockA{\textit{Professorship of Software Engineering} \\
 \textit{University of Technology Chemnitz}\\
 Chemnitz, Germany \\
 siegj@hrz.tu-chemntiz.de}
 }

\maketitle

\begin{abstract}
Background: 
Programming is a fundamental skill in computer science and software engineering specifically. Mastering it is a challenge for novices, which is evidenced by numerous errors that students make during programming assignments.

Objective:
In our study, we want to identify common programming errors in CS2 courses and understand how students evolve over time.

Method:
To this end, we conducted a longitudinal study of errors that students of a CS2 course made in subsequent programming assignments. Specifically, we manually categorized 710 errors based on a modified version of an established error framework.

Result:
We could observe a learning curve of students, such that they start out with only few syntactical errors, but with a high number of semantic errors. During the course, the syntax and semantic errors almost completely vanish, but logical errors remain consistently present.

Conclusion:
Thus, students have only little trouble with learning the programming language, but need more time to understand and express concepts in a programming language.
\end{abstract}


\section{Introduction}
Programming is a fundamental skill for software development. It is also a task with which many students struggle, as evidenced by the high failure rates in programming courses~\cite{Robins2019}. For beginners, it is especially challenging to grasp the linguistic nuances of programming languages, while also learning computational thinking~\cite{Wing06}. This is one reason why it is particularly difficult for students to create programs without syntactic, semantic, or logical errors~\cite{Hristova2003, Brown2017, Mase2022}.

Syntactic programming errors refer to the basic rules of a programming language, such as incorrectly declaring a variable or additional closing parenthesis, while semantic errors go one step further, such as using a variable that has not been declared or incorrectly assigning a floating point value to a variable that can only hold natural numbers~\cite{Linz2011, Hristova2003}. Syntactic and many semantic errors can be detected by compiler error messages~\cite{Jadud2006Compiler, Tabano2011Compiler, DennyCompiler, DyCompiler, Hristova2003}. However, some semantic errors, such as using \texttt{==} for comparison of objects instead of the \texttt{equals} method in Java~\cite{Brown2017}, are not detected on compile time, but still depend on the context of the language.

Logical errors describe that a program does not behave according to its specification, but adheres to the syntactic and semantic rules of a programming language. An example would be a program that is supposed to identify prime numbers, but for some reason identifies 4 as a prime number.~\cite{Hristova2003, Mase2022}

Studies suggest that students still make numerous errors after completing a CS1 course, which even counts for basic syntactic errors~\cite{Kölling2019, EllaSlopiness, Chuang2024, Brown2017}. However, there is only little dedicated research on the specific kinds of errors students make in a CS2 course or how they evolve within one semester. This is a missed opportunity, because it would help educators understand the learning curves of students and how they could help them to overcome the threshold from novice programmer to professional programmer.

With our work, we want to seize that opportunity. To shed more light on student's errors and how they evolve in this particular phase of learning, we replicate a study from McCall and Kölling~\cite{Kölling2019}. Specifically, we use their error categorization framework as a tool to explore and understand student errors. One advantage of this framework is its modular extensibility, which we exploited by adding categories that contain logical errors.

To this end, we analyze student submissions to programming assignments during a CS2 course, and run unit tests on them, so that we can automatically detect logical errors, in addition to compiler errors. We found that students start the course with a high number of semantic errors, and, when faced with new concepts, logical errors as well. This changes during the course, such that semantic errors only play a minor role. Thus, students struggle only at first with learning the semantic rules of the new programming language, and quickly adapt, so that they can focus on implementing a concept, less on how a programming language works.

We make the following contributions:
\begin{itemize}
    \item A replication of the study by McCall and Kölling~\cite{Kölling2019} to describe programming errors in a CS2 course.
    \item The addition of logical error categories to the framework.
    \item A longitudinal view to observe the development of students over the course of a semester.
    \item A publicly available replication package: \url{https://github.com/CSEEandT-25-ErrorPaper/CS2-Error-Categorization}
\end{itemize}

\section{Related Work}
The framework we based our work on was developed by McCall and Kölling based on thematic analysis of student submissions that were automatically collected~\cite{Kölling2014}. In subsequent studies, this framework was evaluated and refined~\cite{Kölling2019}. Thus, it is a well-tested and established framework, making it the best choice for describing errors of students. It consists of 11 categories of errors, of which 4 describe syntactical errors and 6 semantic errors (one additional category contains unspecified errors). Each of the 11 categories is described by several subcategories. For example, the category \emph{9.0 Simple syntactical errors} consists of subcategories, such as \emph{9.8 Extraneous closing curly brace} and \emph{9.3 ; missing}. In its original version, it does not contain categories for logical errors.

Several other frameworks exist, as well. Hristova and others were the first to propose a framework to describe students errors beyond compiler-based analysis~\cite{Hristova2003}. Based on interviews with educators and students, they described the most common errors, which they divide into syntax, semantic, and logical errors, each of which are refined in several subcategories. This framework was tested and extended by Brown and Altamdri as well as Mase and Nel~\cite{Brown2014, Mase2022}. However, it lacks the modular extensibility of the framework developed by McCall and Kölling, due to its flat hierarchy and more rigid categorization. Additionally, while McCall and Kölling's error categorization encompasses all error types identified by Brown and Altamdri, the reverse does not hold true~\cite{Kölling2019, Brown2017}. Their framework has fewer specific subcategories, while the super categories are not meant to contain unspecified programming errors.

There is considerable further research on how to categorize student errors, focusing on logical errors~\cite{Ettles2018}, static analysis errors~\cite{Edwards2017}, or compiler messages\cite{DennyCompiler, Tabano2011Compiler}. However, none of these frameworks are sufficiently detailed or focused on concrete programming errors to describe the variety of student errors that can be observed.

\begin{table*}[!ht]
 \caption{Programming errors per category and task}
    \centering
    \begin{tabular}{lrlrrrrrrrr}
        \toprule
        \textbf{Error category} & \textbf{Error type} & \textbf{Task 1} & \textbf{Task 2}  & \textbf{Task 3} & \textbf{Task 4} & \textbf{Task 5} & \textbf{Task 6} & \textbf{Task 7}  & \textbf{Sum} \\
        \midrule
            \textcolor{gray}{1.0 Incorrect attempt to use variable} & \textcolor{gray}{Semantic} & 
            \textcolor{gray}{5} & \textcolor{gray}{33} & \textcolor{gray}{9} & \textcolor{gray}{0} & \textcolor{gray}{0} & \textcolor{gray}{0} & \textcolor{gray}{1} & \textcolor{gray}{48} \\
            2.0 Incorrect variable declaration & Syntactic &
            2 & 1 & 0 & 0 & 0 & 0 & 0 & 3 \\ 
            \textcolor{gray}{3.0 Incorrect method call} & \textcolor{gray}{Semantic} &
            \textcolor{gray}{1} & \textcolor{gray}{46} & \textcolor{gray}{1} & \textcolor{gray}{0} & \textcolor{gray}{0} & \textcolor{gray}{0} & \textcolor{gray}{0} & \textcolor{gray}{48} \\ 
            4.0 Incorrect method declaration & Syntactic & 
            5 & 15 & 0 & 0 & 0 & 0 & 0 & 20 \\ 
            \textcolor{gray}{5.0 Incorrect constructor call} & \textcolor{gray}{Semantic} &
            \textcolor{gray}{0} & \textcolor{gray}{4} & \textcolor{gray}{0} & \textcolor{gray}{0} & \textcolor{gray}{0} & \textcolor{gray}{0} & \textcolor{gray}{0} & \textcolor{gray}{4} \\ 
            \textcolor{gray}{6.0 Incorrect constructor declaration} & \textcolor{gray}{Semantic} & \textcolor{gray}{0} & \textcolor{gray}{0} & \textcolor{gray}{0} & \textcolor{gray}{0} & \textcolor{gray}{0} & \textcolor{gray}{0} & \textcolor{gray}{0} & \textcolor{gray}{0} \\
            \textcolor{gray}{7.0 Incorrect/attempted use of class or type} & \textcolor{gray}{Semantic} & 
            \textcolor{gray}{0} & \textcolor{gray}{68} & \textcolor{gray}{6} & \textcolor{gray}{5} & \textcolor{gray}{8} & \textcolor{gray}{0} & \textcolor{gray}{2} & \textcolor{gray}{89} \\ 
            \textcolor{gray}{8.0 Semantic error} & \textcolor{gray}{Semantic} & \textcolor{gray}{105} & \textcolor{gray}{66} & \textcolor{gray}{31} & \textcolor{gray}{5} & \textcolor{gray}{1} & \textcolor{gray}{4} & \textcolor{gray}{5} & \textcolor{gray}{217} \\ 
            9.0 Simple syntactical error & Syntactic & 
            7 & 12 & 0 & 4 & 0 & 1 & 0 & 24 \\ 
            10.0 Statement outside method/block &  Syntactic & 
            1 & 1 & 0 & 0 & 0 & 0 & 0 & 2 \\ 
            11.0 Uncategorized & - & 
            1 & 4 & 0 & 0 & 0 & 0 & 0 & 5 \\ 
            12.0 Logical error in working with data structures & Logic &  
            0 & 5 & 83 & 28 & 13 & 11 & 1 & 141 \\
            13.0 Other logical error & Logic &  
            0 & 40 & 22 & 23 & 9 & 1 & 14 & 109 \\
            \addlinespace
            Sum & & 127 & 295 & 152 & 70 & 31 & 17 & 23 & 710 \\
        \bottomrule
    \end{tabular}
    \label{tab:errsPerCategory}
\end{table*}

\section{Study Design}
In this section, we present the study set up and research questions. All material is available at the project's Web site: \url{https://github.com/CSEEandT-25-ErrorPaper/CS2-Error-Categorization}.

\subsection{Research Questions}
Our overarching goal with this work is to understand students' errors and how they evolve during a CS2 course. This way, we can understand the obstacles face to become proficient programmers. To guide our study, we define the following research questions:

\begin{itemize}
    \item RQ$_1$: \RQOne
    \item RQ$_2$: \RQThree
\end{itemize}

The error categories follow the framework by McCall and Kölling~\cite{Kölling2019}. We added two further categories to describe logical errors, with 7 subcategories. Note that the framework contains the category \emph{8.0 Semantic errors} and \emph{9.0 Simple syntactical errors}. To clearly differentiate these categories from the type of errors, we set categories of the framework in \emph{italics}, and when we talk of the broader categories of syntax, semantic, and logical errors, we do not set them in italics. 

\subsection{Participants}
Of all the students enrolled in the CS2 course, we received the permission of 13 to analyze their solutions to mandatory programming assignments. Most have a background in C, Python, or both, from a preceding CS1 course.

\subsection{Tasks}
We defined seven tasks, which covered arrays, classes, lists, binary trees, and graphs. For example, in Task 2, students should implement five classes that interacted with each other to simulate a university in a small program.

The students received a task sheet with descriptions of the classes and methods they needed to implement, which included interfaces for the mandatory classes, and a source code file for each class that already implemented the interface and also provided the constructor declaration. This ensured the compatibility of the submissions with unit tests, which we used to assess the submissions to automatically detect logical errors. Students were encouraged to create additional classes and methods to structure their code.

The students had two weeks to solve each task and had to use the Java API, no third party libraries, but could use code from online or generative AI sources. The students were free to work with an editor of their choice, but we recommended an integrated development environment (IDE), specifically Visual Studio Code. Only compilable code would be graded.

\subsection{Analysis protocol} \label{sec:analysis}
We anonymized the submissions by assigning each student an ID. The first author proceeded to categorize the errors in each submission. If an error did not fit into any of the existing subcategories from \cite{Kölling2019}, we sorted it into an \emph{unspecified} category. If one of these appeared more than two times, we extended the original framework by a new subcategory.

To find the errors, we ran unit tests. If a unit test failed, we documented the cause, fixed it, and ran the tests again. We repeated this process until the submission passed all tests. To not miss any potential errors not captured by the unit tests, we manually evaluated each submission for further errors.

\section{Results}

In total, we evaluated 67 submissions from 13 different students. We excluded one submission (Task 5 from S$_4$) from analysis, since the student ignored the provided interface and strayed too far from the intended task, making meaningful error categorization impossible.

Table~\ref{tab:errsPerCategory} contains an overview of the number of errors per task. Most programming errors occurred in the first three tasks, with Task 2 being the most difficult task with 295 errors.
A considerable drop in the total number of errors occurred after Task 3. This downward trend continued, except for the last task, which has slightly more errors than Task 6.

\subsection{RQ$_1$: \RQOne}
Semantic errors are the most frequent error type. The number of errors was especially high for Tasks 1 and 2, before dropping in Task 3. From Task 4 onwards, they occurred only rarely. The majority of the \emph{8.0 semantic errors} in Task 1 referred to an incorrect use of indexes (63 errors). A regularly occurring \emph{8.0 semantic error} was using comparison operators (\texttt{==}) to compare strings instead of the method \texttt{equals}, which was one of the categories which we added to the framework. 

The most frequent semantic category in Task 2 is \emph{7.0 incorrect/attempted use of a class or type} (68 errors). The errors are mostly related to inheritance or interfaces, such as passing a class instead of an interface as parameter (13 errors), incorrect use of the \texttt{@Override} annotation (11 errors), changing the return type of an inherited method (9 errors), and general interface or inheritance implementation errors (9 errors). 

Notably, students did not struggle with basic syntactic rules, such as \emph{2.0 Incorrect variable declaration}, or \emph{10.0 Statement outside method/block}, each of which occurred only rarely.

Looking at each student over time (cf.\ Table~\ref{tab:student_erroramount}), we observe two patterns: (1) students who continuously made only few errors, such as S$_6$ and S$_9$, and (2), students who made a lot of errors in the beginning and then either improved, such as S$_3$, S$_5$, and S$_8$, or stopped submitting, for example, S$_2$ and S$_{11}$. Notably, the high error count in the first three tasks is caused by few students. In Task 1, three students caused 58\,\% of the errors. This is even more pronounced in Task 2, in which two students made 70\% of errors. And for Task 3, one student caused 63\,\% of the errors. Noticeably, these are never the same students, so each student only had one submission with a high error count.

\greybox{\textbf{Answer RQ1:}
Students seem to struggle mostly with semantic errors, especially when using classes. Basic syntactic structures do not seem to pose problems. Midway through the course, the error rate drops considerably, but some errors persist during the course.
}

\begin{table}[!ht]
    \caption{Programming errors per student. Median in braces denote values without S1, S7, and S11.}
    \centering
    \setlength{\tabcolsep}{3pt}
    \begin{tabular}{lrrrrrrrr}
        \toprule
        \textbf{Student ID} & \multicolumn{7}{c}{\textbf{Task}} & \textbf{Sum} \\
        & 1  & 2  & 3   & 4   & 5   & 6   & 7  &     \\
        \midrule
        S1  & 7  & /   & /   & /   & /   & /   & /  & 7 \\ 
        S2  & 18 & 95  & 16  & /   & /   & /   & / & 129  \\ 
        S3 & 14 & 110 & 20  & 10  & 5   & 3   & 3 & 165  \\ 
        S4 & 2  & 25  & 7   & 10  & n/a   & 2   & 6  & 52 \\
        S5 & 28 & 18  & 5   & 8   & 8   & 2   & 3  & 72 \\ 
        S6 & 2  & 5   & 2   & 8   & 2   & 1   & 4 & 24  \\ 
        S7 & 2  & /   & /   & /   & /   & /   & /  & 2 \\ 
        S8 & 22 & 0   & 3   & 1   & 1   & 0   & 1  & 28 \\ 
        S9 & 0  & 3   & 9   & 2   & 1   & 1   & 0  & 16 \\ 
        S10 & 9  & 12  & 13  & 14  & 4   & 3   & 4  & 59 \\ 
        S11 & 23 & /   & /   & /   & /   & /   & / & 23  \\ 
        S12 & /  & 10  & 66  & 6   & 3   & 2   & 0  & 87 \\ 
        S13 & /  & 17  & 11  & 6   & 7   & 3   & 2 & 46  \\
        \addlinespace
        Median & 9 \textcolor{gray}{(11.5)} & 14.5 & 10 & 8 & 3.5 & 2   & 3 & 46 \textcolor{gray}{(55.5)} \\
        \bottomrule
    \end{tabular}
    \label{tab:student_erroramount}
\end{table}

\begin{table*}[!ht]
 \caption{Logical errors per sub category and task}
    \centering
    \begin{tabular}{lrrrrrrrr}
        \toprule
        \textbf{Logical error} & \textbf{Task 2}  & \textbf{Task 3} & \textbf{Task 4} & \textbf{Task 5} & \textbf{Task 6} & \textbf{Task 7}  & \textbf{Sum} \\
        \midrule
            12.0 Logical error in working with data structures
             &  &  &  &  &  &  &  \\
            
            12.1 ... in working with classes 
            & 5 & 42 & 5 & 0 & 0 & 0 & 52 \\
            12.3 ... in working with custom implemented linked lists 
            & 0 & 41 & 23 & 0 & 0 & 0 & 64  \\ 
            12.5 ... in working with a queue or a stack 
            & 0 & 0 & 0 & 2 & 0 & 0 & 2  \\
            12.2 ... in working with graphs
            & 0 & 0 & 0 & 0 & 11 & 1 & 12 \\ 
            12.2.1 ... in working with a binary tree 
            & 0 & 0 & 0 & 11 & 0 & 0 & 11  \\ 
            13.0 Other logical error
             &  &  &  &  &  &  &  \\
            13.1 Incomplete logic in a function 
            & 11 & 10 & 14 & 2 & 0 & 12 & 49  \\ 
            13.2 Unspecified logical error 
            & 29 & 12 & 9 & 7 & 1 & 2 & 60  \\ 
            \addlinespace
            Sum & 45 & 105 & 51 & 22 & 12 & 15 & 250 \\
        \bottomrule
    \end{tabular}
    \label{tab:errsLogicCategory}
\end{table*}

\subsection{RQ$_2$: \RQThree}

Logical errors started appearing in Task 2. We show an overview of the logical errors in Table~\ref{tab:errsLogicCategory}. Overall, we observed 7 subcategories of logical errors. In Task 3, students made the most logical errors (105 errors). In the subsequent tasks, the amount of logical errors decreased by about half each task until Task 6 with 12 errors. There is no further drop to Task 7, but a slight increase, with most of the 15 errors relating to incomplete implementation of logic (12 errors). Starting from Task 3, logical errors were consistently the most frequent ones. 

The majority of errors relate to the concept that was part of a task. Linked lists especially caused troubles for students, followed by working with classes. Typically, the amount of programming errors decreases the second time a concept was part of the implementation. One exception were classes, which caused the most trouble in the second task they occurred in.

\newpage
\greybox{\textbf{Answer RQ2:}
Students tend to make logical errors when implementing new data structures or concepts, but these errors decrease with repeated practice.
}

\section{Discussion}
\subsection{Learning Curve}
The learning curve we observed can be summarized as follows: Students face only few problems when learning basic Java syntax. Instead, they start with a high rate of semantic errors, which relate to concepts that have no explicit syntactic equivalent in C (e.g., structs vs.\ classes) or behave a bit differently than in Java (e.g., type casting). This takes about the first half of the course, and, while students begin to make fewer semantic errors, logical errors are increasing. This learning curve is in line with studies demonstrating that students transfer their knowledge from one programming language to another~\cite{tshukudu2020understanding, tshukudu2021pedagogy}. 

In other words, students quickly capture basic syntactic rules of Java, based on their knowledge of C, which they learned in the preceding CS1 course. Basic rules can be directly transferred from C to Java, such as variable declaration or that statements have to be within one block. Since these basic rules pose no troubles, students focus on learning how to transfer concepts that they already know, such as structs and static types, to how these concepts are expressed in Java. This takes some effort, because these concepts do not have the same syntactical representation. Especially the first tasks that introduce classes pose difficulties for students. In Task 2, the focus is on the newly introduced concept of classes, and we observe a peak of errors in the \emph{incorrect/attempted use of class or type} category. This ripples through to Task 3, in which \emph{logical errors in working with classes} occur most often. But from Task 4, there are no errors related to classes. Thus, once students grasped that structs and classes are comparable, they do not have to learn something entirely new, but learn how to express something that they already know in a new programming language. Then, they can start focusing on understanding and implementing new concepts.

Such an interpretation is supported by the development of the logical errors, which are the most frequent ones starting from Task 3. From that task on, students had to work with a new data structure every time, and for each task, the logical errors are related to the concrete data structure. Other studies have demonstrated similar peaks in errors around certain newly introduced concepts~\cite{bryce2010cs1cs2, Mase2022}, possibly hinting at them being threshold concepts, which are difficult to learn, but once grasped, cannot be unlearned\cite{Boustedt2007, Sanders2016}. Concepts of object orientation and data structures have also been suggested as threshold concepts~\cite{Mostroem2008}. Thus, after having expressed data structures and related algorithms in Java, students may have overcome the threshold concept.


\subsection{Nudging Students to Submit Compilable Code}
The setup of the course explicitly fostered students submitting working code, thus, avoiding errors that can be detected by a compiler. First, we explicitly encouraged students to implement the assignments with an IDE, which alerts students of such errors during programming, supporting them in spotting and possibly also fixing errors based on the compiler error messages. Although compiler error messages are often confusing for students~\cite{Becker2019, Hristova2003}, they might have familiarized themselves in the previous CS1 course with the interpretation of compiler error messages to fix errors. Second, our submission procedure supported and encouraged students to submit solutions that compile, due to the templates we provided. Thus, with dedicated support to learn the syntactic rules of a new programming language, we can help students focus on learning the intended concepts and data structures and how to express related problems in a programming language, thereby fostering computational thinking~\cite{Wing06}.

\subsection{Student Development}

Some students made a high number of errors in the beginning, but then either improved or stopped submitting. Interestingly, it was never the same students with the high error count, so all students improved quickly. This might indicate that the feedback they receive from one error-loaded submission leads to them paying more attention to details and produce more correct code. The students who considerably improved might have mastered one or two threshold concepts, lifting them to a higher level of programming skill.

All students who finished the course did so with low error counts in their final submissions, suggesting that it is important for students to push through these early obstacles, as it is still possible to learn and improve considerably. This is interesting to interpret in the context of the work on predicting student success in programming courses, which found that students who fall behind early in programming courses tend to stay behind~\cite{ahadi_falling_2014}. These studies focused on CS1 students, and showed that, if students fail to grasp basic concepts, such as how variables work, they have trouble catching up. For our CS2 course, we could not observe such a pattern, since none of the students stopped submitting with high error rates. Instead, we interpret that motivation is an important driving factor to succeed in a programming course, which is also in line with previous studies~\cite{lishinski2019motivation, 
schantong2024}. Possibly, the students could have also learned from the preceding course that pushing through obstacles leads to eventual success. 

\begin{table}[!ht]
 \caption{Percentage of errors per category across different studies}
 \centering
 \begin{tabular}{p{4.6cm}p{1.32cm}p{1.32cm}l}
 \toprule
 \textbf{Error category} & \textbf{McCall and Kölling~\cite{Kölling2019}} & \textbf{Current Study} \\
 \midrule
 	\textcolor{gray}{1.0 Incorrect attempt to use variable}  &
 	\textcolor{gray}{19.9\,\%}  & \textcolor{gray}{6.8\,\%} \\
 	2.0 Incorrect variable declaration  &
 	5.2\,\% & 0.4\,\%	\\
 	\textcolor{gray}{3.0 Incorrect method call} &
 	\textcolor{gray}{18\,\%}  & \textcolor{gray}{6.8\,\%}         	\\
 	4.0 Incorrect method declaration &
 	7.9\,\% & 2.8\,\%    	\\
 	\textcolor{gray}{5.0 Incorrect constructor call} &
 	\textcolor{gray}{0.8\,\%} & \textcolor{gray}{0.6\,\%} 	\\
 	\textcolor{gray}{6.0 Incorrect constructor declaration} &
 	\textcolor{gray}{0.4\,\%} & \textcolor{gray}{0\,\%} \\
 	\textcolor{gray}{7.0 Incorrect/attempted use of class or type} &
 	\textcolor{gray}{6\,\%}  & \textcolor{gray}{12.5\,\%}	\\
 	\textcolor{gray}{8.0 Semantic error} &
 	\textcolor{gray}{12.7\,\%}  & \textcolor{gray}{30.6\,\%}          	\\
 	9.0 Simple syntactical error &
 	20.5\,\%  & 3.4\,\%              	\\
 	10.0 Statement outside method/block &
 	0.2\,\% & 0.3\,\%       	\\
 	11.0 Uncategorized 	&
 	8.5\,\%  & 0.7\,\%                      	\\  
 	12.0 Logical error in working with data structures &
    n/a  & 19.9\,\%                      	\\  
    13.0 Other logical error &
 	n/a & 15.4\,\%  	\\
 	\addlinespace
    Sum syntax errors & (34\,\%) 373 &(8\,\%) 55\\
    \textcolor{gray}{Sum semantic errors} & \textcolor{gray}{(58\,\%) 638} & \textcolor{gray}{(58\,\%) 408}\\    
    Sum logic errors & n/a &(35\,\%) 250\\
    \addlinespace
 	Amount of investigated errors & 1105 & 710 \\
\bottomrule
\end{tabular}
\label{tab:errorcomparsion}
\end{table}

\subsection{Comparison to McCall and Kölling~\cite{Kölling2019}}

To set our results into context, we compare the errors that we found with the ones from McCall and Kölling in Table~\ref{tab:errorcomparsion} (the comparison with other frameworks is available on the project's Web site). The original framework does not contain logical errors, likely because the authors did not automatically detect them~\cite{Kölling2019}. Our analysis procedure based on automatic unit testing allowed us to also detect and categorize logical errors. Thus, for the comparison, we only focus on syntax and semantic errors. We found 55 \emph{syntax errors}, while in the original work, there were 373 syntax errors. Especially the 21\,\% \emph{9.0 Simple syntactical errors} are a difference, as we did observe only few of them (3\,\%). Regarding semantic errors, we observed 408, in contrast to 638 in the original study. The most noteworthy difference is the category of \emph{8.0 Semantic errors}, in which we found 31\,\%, compared to 13\,\% in the original study. Thus, our CS2 students seem to be further along in their learning process, compared to the students from McCall and Kölling. Unfortunately, the Blackbox data prohibits us from directly comparing their results to ours. Given that, in the original study, the Blackbox data set is a snapshot sometime during the process of programming, it gives a more general overview of types of errors, while in our sample, we observed the last snapshot of a submitted solution. Thus, our data points more toward the potential of students to provide an error-free, compilable program, while in the original framework, it gives information about all the types of errors during the process. Thus, our data hints more toward the difficulties that students cannot overcome themselves and that might be especially important to address in CS2 courses to support them on their way to become professional programmers.

\section{Threats to Validity}

We selected the most suitable error categorization for our data set. Although a different framework may have produced a different outcome, thus posing a threat to \emph{construct validity}, this framework nevertheless gives a useful overview of students' programming errors and how they evolve.

Our submission procedure encouraged students to provide compilable solutions, which might threaten \emph{internal validity}, because it might hide some of the more basic errors that students may have made during programming. With a different procedure, for example, including regular snapshots, we likely would have observed a different distribution of errors. Nevertheless, our data points toward the potential of CS2 students to avoid errors.

Unfortunately, our sample was rather small, so it is difficult to generalize our findings, which threatens \emph{external validity}. Nevertheless, we reviewed a large body of code for different tasks and over the course of an entire semester, giving us valuable insights into the evolution of students.

\section{Conclusion}
In general, we could replicate the errors that students make during programming according to the framework by McCall and Kölling, but with some nuances. Specifically, the concrete data structures and related algorithms seem to be a struggle for students in a CS2 course, which may hint at the data structures being threshold concepts. Semantic errors seem to be a hindrance at the beginning of the course that students overcome with some time. Syntactic errors, which are considered typical for novice programmers, do not seem to pose a problem for CS2 students.


\section*{Data Availability}
All data and supplementary material is available at: \newline
\url{https://github.com/CSEEandT-25-ErrorPaper/CS2-Error-Categorization}

\bibliographystyle{IEEEtran}
\bibliography{bibliography}

\end{document}